\documentclass{article}

\usepackage{PRIMEarxiv}

\usepackage[utf8]{inputenc} 
\usepackage[T1]{fontenc}    
\usepackage{hyperref}       
\usepackage{url}            
\usepackage{booktabs}       
\usepackage{amsfonts}       
\usepackage{nicefrac}       
\usepackage{microtype}      
\usepackage{lipsum}
\usepackage{fancyhdr}       
\usepackage{graphicx}       
\graphicspath{{media/}}     

\usepackage{bm}
\usepackage{adjustbox}

\title{GreenBIQA: A Lightweight Blind Image Quality Assessment Method}

\author{
  Zhanxuan Mei \\
  University of Southern California \\
  Los Angeles, USA\\
  \texttt{zhanxuan@usc.edu} \\
   \And
  Yun-Cheng Wang \\
  University of Southern California \\
  Los Angeles, USA\\
  \texttt{yunchenw@@usc.edu} \\
   \And
  Xingze He \\
  Meta Platform, Inc. \\
  Menlo Park, California, USA\\
  \texttt{xingze.he@fb.com} \\
   \And
  C.-C. Jay Kuo \\
  University of Southern California \\
  Los Angeles, USA\\
  \texttt{cckuo@sipi.usc.edu} \\
}

\begin{document}
\maketitle

\begin{abstract}
Deep neural networks (DNNs) achieve great success in blind image quality assessment (BIQA) with large pre-trained models in recent years. Their solutions cannot be easily deployed at mobile or edge devices, and a lightweight solution is desired. In this work, we propose a novel BIQA model, called GreenBIQA, that aims at high performance, low computational complexity and a small model size. GreenBIQA adopts an unsupervised feature generation method and a supervised feature selection method to extract quality-aware features. Then, it trains an XGBoost regressor to predict quality scores of test images. We conduct experiments on four popular IQA datasets, which include two synthetic-distortion and two authentic-distortion datasets. Experimental results show that GreenBIQA is competitive in performance against state-of-the-art DNNs with lower complexity and smaller model sizes. 
\end{abstract}


\section{Introduction}\label{sec:introduction}

Image quality assessment (IQA) aims to evaluate image quality at various
stages of image processing such as image acquisition, transmission, and
compression.  Based on the availability of undistorted reference images,
objective IQA can be classified into three categories
\cite{lin2011perceptual}: full-reference (FR), reduced-referenced (RR)
and no-reference (NR). The last one is also known as blind IQA (BIQA).
FR-IQA metrics have achieved high consistency with human subjective
evaluation. Many FR-IQA methods have been well developed in the last two
decades such as SSIM \cite{wang2004image} and FSIM \cite{zhang2011fsim}.
RR-IQA metrics only utilize features of reference images for quality
evaluation. In some application scenarios (e.g., image receivers), users
cannot access reference images so that NR-IQA is the only choice. BIQA
methods attract growing attention in recent years. 

Generally speaking, conventional BIQA methods consist of two stages: 1)
extraction of quality-aware features and 2) adoption of a regression
model for quality score prediction.  As the amount of user generated
images grows rapidly, a handcrafted feature extraction method is limited
in its power of modeling a wide range of image content and distortion
characteristics.  Recently, due to the strong feature representation
capability of deep neural networks (DNNs), DNN-based BIQA methods have
been proposed and they have achieved significant performance improvement
\cite{yang2019survey}. Yet, collecting large-scale IQA datasets with
user annotations is expensive and time-consuming. Moreover, for IQA
datasets of a limited size, DNN-based BIQA methods tend to overfit the
training data and do not perform well against the test data.  To address
this problem, DNN-based solutions often rely on a huge pre-trained
network which is trained by a large dataset. The large model size
demands high computational complexity and memory requirement. Such a
solution cannot be easily deployed at mobile or edge devices, where a
lightweight solution is essential. Here, we propose a lightweight BIQA
method called GreenBIQA. It aims to achieve high performance that is
competitive with DNN-based solutions yet demands much less computing
power and memory.  To this end, for a video source, GreenBIQA can
predict perceptual quality scores frame by frame in real time.

To enlarge the training sample number in GreenBIQA, we first crop
distortion-sensitive patches from images. Then, an unsupervised
representation determination method is used to obtain a large number of joint
spatial-spectral features from each patch. Next, we adopt a supervised
feature selection method, called the relevant feature test (RFT)
\cite{yang2022supervised}, to reduce the dimension of quality-aware
features.  Finally, an XGBoost \cite{chen2016xgboost} regressor is used
to predict the final quality scores. 

There are two main contributions of this work.
\begin{itemize}
\item A novel, lightweight and modularized BIQA method is proposed.  The 
components of GreenBIQA include image augmentation, unsupervised feature determination, supervised quality-aware feature selection, and MOS prediction. It can
extract quality-aware features efficiently and yield accurate quality
predictions. 
\item We conduct experiments on four widely-used BIQA datasets to
demonstrate the advantages of GreenBIQA.  It outperforms all
conventional BIQA methods in prediction accuracy. Furthermore, its
prediction performance is highly competitive with state-of-the-art
DNN-based methods, which is achieved with a significantly smaller model
size and at a much faster speed. 
\end{itemize} 
The rest of this paper is organized as follows. Related work is
reviewed in Sec. \ref{sec:related}. GreenBIQA is proposed in 
Sec. \ref{sec:method}. Experimental results are shown in Sec.
\ref{sec:experiments}. Finally, concluding remarks are given in
Sec. \ref{sec:conclusion}.

\section{Related Work}\label{sec:related}
Quite a few BIQA methods have been proposed in the last two decades.
Existing work can be classified into conventional and DNN-based two
categories as reviewed below.

\begin{figure*}[!htbp]
\centering
\includegraphics[width=1.0\linewidth]{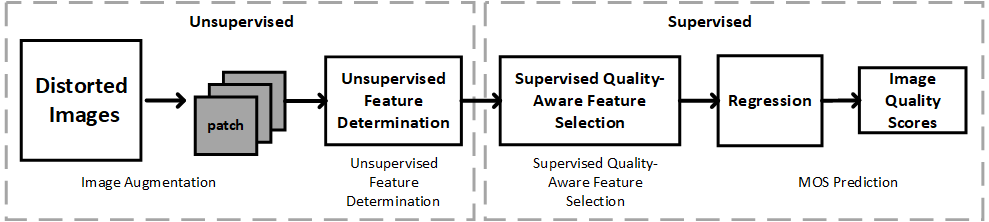}\\
\caption{An overview of the proposed GreenBIQA method.}\label{fig:GreenIQA-pipeline}
\end{figure*}


\subsection{Conventional BIQA Method}

Conventional BIQA methods extract quality-aware features from input
images. Then, they train a regression model, e.g., Support Vector
Regression (SVR) \cite{awad2015support} or XGBoost
\cite{chen2016xgboost}, to predict the quality score based on these
features.  One representative class of conventional methods relies on
the Natural Scene Statistics (NSS). For example, BIQI
\cite{mittal2012making} developed the Distorted Image Statistics (DIS)
to capture the NSS changes caused by various distortions. Then, it
adopted a distortion-specific quality assessment framework.  By
following a similar two-stage framework, DIIVINE \cite{moorthy2011blind}
improved the perceptual quality prediction performance furthermore.
Instead of computing distortion-specific features, BRISQUE
\cite{mittal2012no} used scene statistics to quantify loss of
naturalness caused by distortions.  It operated in the spatial domain
with low complexity. Although NSS-based methods are simple, their
features are not powerful enough to handle a wide range of distortion
types. 

Another class of conventional methods leverage the use of codebooks.
They share one common framework; namely, local feature extraction,
codebook construction, feature encoding, spatial pooling, and quality
regression. For instance, CORNIA \cite{ye2012unsupervised} extracted
raw-image-patches from unlabeled images and used clustering to build a
dictionary. Then, an image can be represented by soft-assignment coding
with spatial pooling of the dictionary. Finally, SVR was adopted to map
the encoded quality-aware features to quality scores. By following the
same structure, HOSA \cite{xu2016blind} was developed to improve the
performance and reduce the computational cost.  Besides the mean of each
cluster, some high order statistical information (e.g., dimension-wise
variance and skewness) of clusters can also be aggregated to form a
small codebook. Codebook-based methods demand high-dimensional
handcrafted feature vectors. Again, they cannot handle a wide range of
distortion types effectively.

\subsection{Deep-learning Based BIQA Method}

Deep-learning-based methods have been investigated to solve the BIQA
problem. A BIQA method based on the convolutional neural network (CNN),
consisting of one convolutional layer with max and min pooling and two
fully connected layers was proposed in \cite{kang2014convolutional}.
To alleviate accuracy discrepancy between FR-IQA and NR-IQA, BIECON
\cite{kim2016fully} used FR-IQA to compute proxy quality scores for
image patches and, then, used them to train the neural network. WaDIQaM
\cite{bosse2017deep} adopted a deep CNN model that contains ten
convolutional layers and two fully connected layers.  

Yet, early networks can only handle synthetic distortions (i.e., with
given distortion types generated by humans). To achieve better
performance, recent work adopts a DNN pre-trained by large datasets. For
example, a DNN was pre-tained by the ImageNet \cite{deng2009imagenet}
and synthetic-distortion datasets in DBCNN \cite{zhang2018blind}.
Furthermore, to reflect diverging subjective perception evaluation on an
image, PQR \cite{zeng2018blind} adopted a score distribution to represent
image quality. A hyper network, which adjusts the quality prediction
parameters adaptively, was proposed in HyperIQA \cite{su2020blindly}.
Both local distortion features and global semantic features were
aggregated in this work to gather fine-grained details and holistic
information, respectively. Afterwards, image quality was predicted based
on the multi-scale representation. Despite the good performance of
pre-trained DNNs, their huge model sizes are difficult to deploy in many
real-world applications.

\section{GreenBIQA Method}\label{sec:method}

As shown in Fig. \ref{fig:GreenIQA-pipeline}, GreenBIQA adopts a
modularized solution that consists of the following four modules: (1)
image augmentation, (2) unsupervised feature determination, (3)
supervised quality-aware feature selection, and (4) MOS prediction. They
are elaborated below. 

\begin{figure}[!h]
\centering
\includegraphics[width=0.7\linewidth]{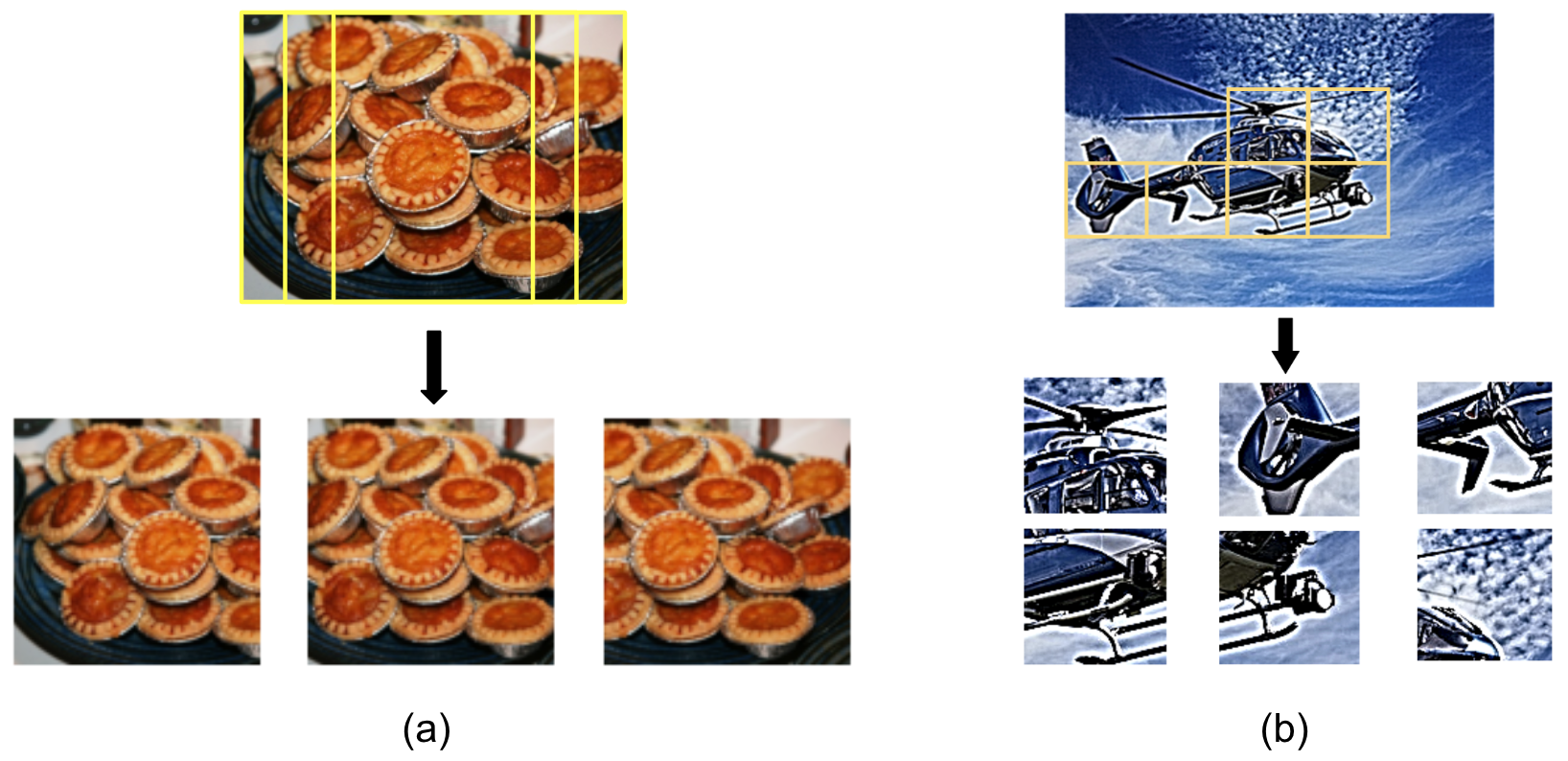}\\
\caption{An example of cropped subimages for (a) authentic-distortion and
(b) synthetic-distortion datasets.}\label{fig:cropping}
\end{figure}

\subsection{Image Augmentation} 
Image augmentation is implemented to enlarge
the number of training samples. This is achieved by image flipping and
cropping. All obtained subimages are assigned the same MOS as their
source image. To ensure the overall perceptual quality of a subimage, we adopt two different image augmentation strategies for BIQA
datasets for authentic-distortions and synthetic-distortions, respectively.

For authentic-distortion datasets such as KonIQ-10K
\cite{hosu2020koniq}, distortions are introduced during capturing.
Collected images have a diverse type and a wide range of distortions.
Besides, distortions are not uniformly applied to images. To preserve
the perceptual quality of subimages and avoid loss of semantic and/or
content information, we crop larger subimages out of source images as
shown in Fig. \ref{fig:cropping} (a). The cropped subimages can be
overlapped. Each subimage is a squared one. Its size is determined by
the height of the source image. We select the left-most, center and
right-most three subimages. 

In contrast, for synthetic-distortion datasets such as KADID-10K
\cite{lin2019kadid}, all distortions are added by humans. They are often
applied to original undistorted images uniformly with few exceptions
(e.g., color distortion in localized regions in KADID-10K).
Consequently, we can crop out patches of smaller sizes and treat them as
subimages to get even more training samples. An example of
non-overlapping cropped patches from an image in KADID-10K is shown in
Fig. \ref{fig:cropping} (b). We select patches that have higher
frequency components. Thus, patches containing foreground objects are
more likely to be selected than those containing simple background. We
feed selected patches to the next module. 

\begin{figure}[!h]
\centering
\includegraphics[width=0.7\linewidth]{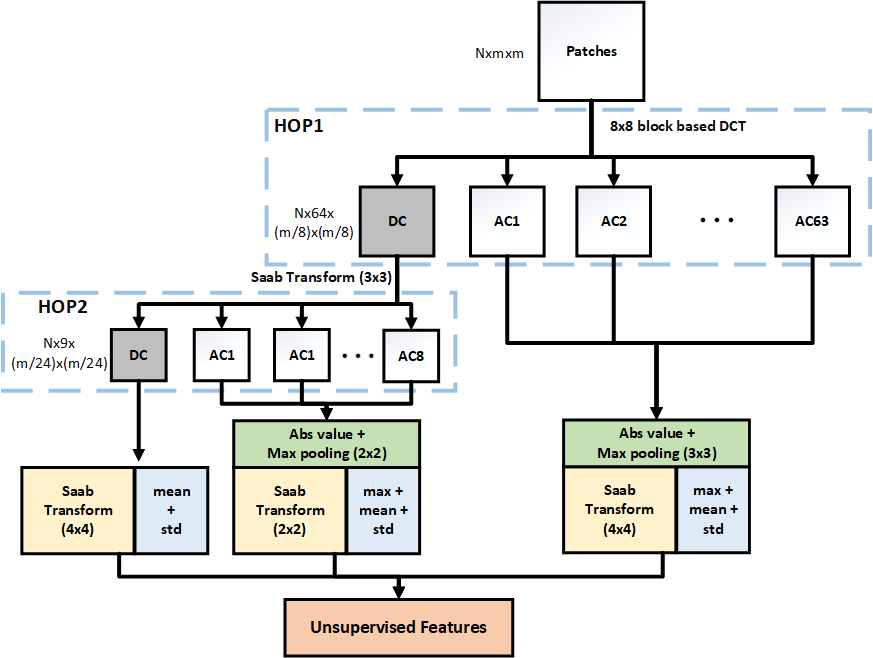}\\
\caption{The block diagram of unsupervised feature determination.}\label{fig:diagram}
\end{figure}


\subsection{Unsupervised Feature Determination} 
Given subimages obtained
above, we derive a set of features to represent the subimages in an
unsupervised manner using a two-hop hierarchy as shown in Fig.
\ref{fig:diagram}, where HOP1 and HOP2 are used to capture local and
global representations, respectively. Since image and video coding
standards all adopt block-based DCT, the processing of GreenBIQA also
operates in the DCT domain for faster speed. 

In HOP1, we partition an input subimage into non-overlapping blocks of
size $8 \times 8$ and compute the DCT coefficients of each block. It is
done for the Y, U and V channels separately. This step can be skipped if
the quality assessment software and the image decoding software can be
integrated since the DCT coefficients can be directly obtained from
compressed images. DCT coefficients of each block are scanned in the
zigzag order, leading to one DC coefficient and 63 AC coefficients,
denoted by AC1-AC63. Since the DC coefficients of spatially adjacent
blocks are correlated, we apply the Saab transform
\cite{kuo2019interpretable} to decorrelate them in HOP2. Specifically,
we partition DC coefficients from HOP1 into non-overlapping blocks of
size $3 \times 3$. To implement the Saab transform, nine DC
coefficients are flattened into a 9-D vector denoted by ${\bf y}=(y_1,
\cdots, y_9)^T$. The mean of the nine DC coefficients,
$\bar{y}=(\sum_{i=1}^9 y_i)/9$, defines the DC in HOP2.  Next, the
principle component analysis (PCA) is applied to the mean-removed vector
${\bf y'}={\bf y}-\bar{y}(1, \cdots, 1)^T$ to yield eight AC
coefficients, denoted by AC1 to AC8 in HOP2. 

There are 63 AC coefficients in each block of HOP1 and 1DC and 8 AC
coefficients in each block of HOP2. It is essential to aggregate them
spatially to reduce the feature number. To lower the feature dimension,
we first take their absolute values and conduct maximum pooling.
Next, we adopt the following operations to generate two sets of features.
\begin{itemize}
\item Compute the maximum value, the mean value, and the standard deviation 
of the same coefficient across the spatial domain. 
\item Conduct the Saab transform on spatially adjacent regions for
further dimension reduction and use the Saab coefficients as spectral features.
\end{itemize}
The above two sets of features are concatenated to form a set of unsupervised features. 

\begin{figure}[!htbp]
\centering
\includegraphics[width=0.5\linewidth]{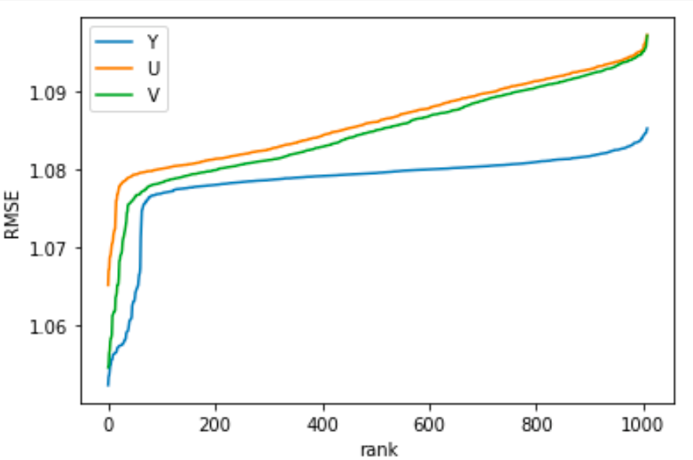}\\
\caption{RFT results in Y, U and V three channels.}\label{fig:ranking}
\end{figure}

\subsection{Supervised Quality-Aware Feature Selection}
To reduce the feature
dimension furthermore, we select quality-aware features from the set of
unsupervised features obtained in the previous module. This is
accomplished by a recently developed method called the relevant feature
test (RFT) \cite{yang2022supervised}. For each 1D feature, RFT splits
its dynamic range into two subintervals with a certain partitioning
point and uses training samples to calculate a cost function. In our
current context, we choose the weighted average of root-mean-squared errors 
(RMSE) of the two subintervals as the cost function.  The set of features 
with the smallest cost functions are selected as quality-aware features. 
We order features of the smallest to the largest cost functions for
Y, U, V three channels in Fig. \ref{fig:ranking}. We see clearly an
elbow point in all three curves. We treat feature dimensions before the
elbow point as quality-aware features. All selected feature dimensions
are concatenated to form the set of quality-aware features.

\subsection{MOS prediction} After quality-aware features are selected, we
adopt the XGBoost \cite{chen2016xgboost} regressor as the quality score
prediction model that maps $d$-dimensional quality-aware features to a
single quality score. Since we extract $k$ subimages from each image in
image augmentation, we have $k$ predicted MOS values for each distorted
image. Finally, we adopt a mean filter to ensemble the predictions from
all subimages to yield the final MOS prediction for a test distorted image. 

\section{Experiments}\label{sec:experiments}

\begin{table}[t]
\centering
\caption{Statistics of IQA datasets.}
\begin{tabular}{ l | cccc }
\hline
Dataset & Dist. & Ref. & Dist. Types &Scenario \\
\hline
CSIQ \cite{larson2010most}      &866 &30 &6 &Synthetic\\
KADID-10K \cite{lin2019kadid}      &10,125 &81 &25 &Synthetic\\
\hline
LIVE-C \cite{ghadiyaram2015massive}   &1,169 &N/A &N/A &Authentic\\
KonIQ-10K \cite{hosu2020koniq}   &10,073 &N/A &N/A &Authentic\\
\hline
\end{tabular}
\label{table:dataset}
\end{table}

\subsection{Datasets} 
We evaluate GreenBIQA on four IQA datasets, including
two synthetic datasets and two authentic datasets. Their statistics are
given in Table \ref{table:dataset}.  CSIQ \cite{larson2010most} and
KADID-10K \cite{lin2019kadid} are two synthetic-distortion datasets,
where multiple distortions of various levels are applied to a set of
reference images. LIVE-C \cite{ghadiyaram2015massive} and KonIQ-10K
\cite{hosu2020koniq} are two authentic-distortion datasets, which
contains a diverse range of distorted images captured by various cameras
in the real world. 

\subsection{GreenBIQA Training}
We crop 25 patches of size $96\times96$ and 6
subimages of size $384\times384$ from images in synthetic- and
authentic-distortion datasets, respectively. The number of quality-aware
features for each of the Y, U, V channels is set as 2,500, 600, and 600 for
authentic and synthetic images, respectively. For the XGBoost regressor,
the learning rate is set to 0.05 and the max\_depth and
number\_of\_estimators are set to 5 and 1000, respectively. We adopt the
early stopping for the XGBoost regressors. All the experiments are run
on a server with Intel(R) Xeon(R) E5-2620 CPU. 

\subsection{Evaluation Metrics} The performance is measured by the Pearson
Linear Correlation Coefficient (PLCC) and the Spearman Rank Order
Correlation Coefficient (SROCC). PLCC is used to measure the
linear correlation between predicted scores and subjective
quality scores. It is defined as
\begin{equation}
\textit{PLCC} = 1 - \frac{\sum_{i}(p_i - p_m)(\hat{p_i}-\hat{p_m})}
{\sqrt{\sum_{i}(p_i - p_m)^2}\sqrt{\sum_{i}\hat{p_i}-\hat{p_m})^2}},
\end{equation}
where $p_i$ and $\hat{p_i}$ denote the predicted score and the
subjective quality score, $p_m$ and $\hat{p_m}$ denote the mean of
predicted score and the subjective quality score, respectively. SROCC is
adopted to measure the monotonicity between predicted and subjective
quality scores. It is defined as
\begin{equation}
\textit{SROCC} = 1 -  \frac{6\sum_{i=1}^{L}(m_i - n_i)^2} {L(L^2-1)},
\end{equation}
where $m_i$ is the rank of $p_i$ in the predicted scores, $n_i$ is the
rank of $\hat{p_i}$ in the subjective quality score and $L$ is the
number of images. We adopt the standard evaluation procedure by
splitting each dataset into 80\% for training and 20\% for testing.
Furthermore, 10\% of training data is used for validation. We run
experiments 10 times and report the median PLCC and SROCC values. For
synthetic-distortion datasets, splitting is implemented based on
reference images to avoid content overlapping. 

\begin{table*}[!htbp]
\centering
\caption{Results on multiple IQA databases.}
\begin{tabular}{ l  cc cc cc cc  c }
\toprule
BIQA Method & \multicolumn{2}{c}{CSIQ} & \multicolumn{2}{c}{LIVE-C} & \multicolumn{2}{c}{KADID-10K} & \multicolumn{2}{c}{KonIQ-10K} & Model Size \\
\cmidrule(l){2-3} \cmidrule(l){4-5} \cmidrule(l){6-7} \cmidrule(l){8-9}
 & SROCC & PLCC & SROCC & PLCC & SROCC & PLCC & SROCC & PLCC & (MB) \\ \midrule
NIQE \cite{mittal2012making}     &0.627 &0.712 &0.455 &0.483 &0.374 &0.428 &0.531 &0.538 &-\\
BRISQUE \cite{mittal2012no}     &0.746 &0.829 &0.608 &0.629 &0.528 &0.567 &0.665 &0.681 &-\\ \midrule
CORNIA \cite{ye2012unsupervised}&0.678 &0.776 &0.632 &0.661 &0.516 &0.558 &0.780 &0.795 &7.4\\
HOSA \cite{xu2016blind}         &0.741 &0.823 &0.661 &0.675 &0.618 &0.653 &0.805 &0.813 &0.23\\ \midrule
BIECON \cite{kim2016fully}      &0.815 &0.823 &0.595 &0.613 &-     &-     &0.618 &0.651 &35.2\\
WaDIQaM \cite{bosse2017deep}    &\textbf{0.955} &\textbf{0.973} &0.671 &0.680 &-     &-     &0.797 &0.805 &25.2\\ \midrule
PQR \cite{zeng2018blind}        &0.872 &0.901 &0.857 &0.882 &-     &-     &0.880 &0.884 &235.9\\
DBCNN \cite{zhang2018blind}     &0.946 &0.959 &0.851 &0.869 &0.851 &\textbf{0.856} &0.875 &0.884 &54.6\\
HyperIQA \cite{su2020blindly}   &0.923 &0.942 &\textbf{0.859} &\textbf{0.882} &\textbf{0.852} &0.845 &\textbf{0.906} &\textbf{0.917} &104.7\\ \midrule
GreenBIQA (Ours) &0.925 &0.936 &0.673 &0.689 &0.847 &0.848 &0.812 &0.834 &1.9\\ \bottomrule
\end{tabular}
\label{table:biqa}
\end{table*}

\begin{table}[!htbp]
\centering
\caption{SROCC of individual distortion types in CSIQ.}
\begin{tabular}{ l | cc cc cc}
\hline
& WN & JPEG & JP2K & FN & GB & CC \\ \hline
BRISQUE \cite{mittal2012no}  &0.723 &0.806 &0.840 &0.378 &0.820 &0.804 \\ 
HOSA \cite{xu2016blind}     &0.604 &0.733 &0.818 &0.500 &0.841 &0.716 \\ 
WaDIQaM \cite{bosse2017deep}  &\textbf{0.974} &0.863 &0.947 &0.882 &\textbf{0.979} &\textbf{0.923} \\ 
HyperIQA \cite{su2020blindly} &0.927 &0.934 &\textbf{0.960} &\textbf{0.931} &0.915 &0.874 \\ \hline
GreenBIQA (Ours)     &0.937 &\textbf{0.971} &0.947 &0.928 &0.922 &0.829 \\ \hline
\end{tabular}
\label{table:distortion}
\end{table}

\subsection{Performance Benchmarking} 
We compare the performance of GreenBIQA
with four conventional and five deep-learning-based methods, and report
their results in Table \ref{table:biqa}. We divide them into four
categories.  
\begin{itemize}
\item NIQE \cite{mittal2012making} and BRISQUE \cite{mittal2012no} use
NSS-based handcrafted features. 
\item CORNIA \cite{ye2012unsupervised} and HOSA \cite{xu2016blind} are
codebook-learning methods. 
\item BIECON \cite{kim2016fully} and WaDIQaM \cite{bosse2017deep} are
deep-learning methods without pre-training. 
\item PQR \cite{zeng2018blind}, DBCNN \cite{zhang2018blind}, and
HyperIQA \cite{su2020blindly} are deep-learning methods with
pre-training. 
\end{itemize}

As shown in Table \ref{table:biqa}, GreenBIQA outperforms conventional
BIQA methods in all four datasets. It also outperforms two
deep-learning methods without pre-training in both authentic-distortion
datasets. As compared with deep-learning methods with pretraining,
GreenBIQA achieves competitive or even better performance in
synthetic-distortion datasets (i.e., CSIQ and KADID-10K). Furthermore,
we show the SROCC performance of individual distortions in the CSIQ
dataset in Table \ref{table:distortion}. There are six distortion types
in CSIQ. The best performer in each category is selected for
comparison. GreenBIQA performs especially well for JPEG distortion. This
could be attributed to the fact that its features are extracted in the
DCT domain. For authentic-distortion datasets, GreenBIQA performs
better than traditional deep-learning methods such as BIECON and
WaDIQaM, which demonstrates that it can generalize well to diverse
distortions. There is a performance gap between GreenBIQA and
deep-learning models with pre-training. Yet, these pre-trained models
have much larger model sizes as a tradeoff. 

\begin{figure*}[!htbp]
\centering
\includegraphics[width=1\linewidth]{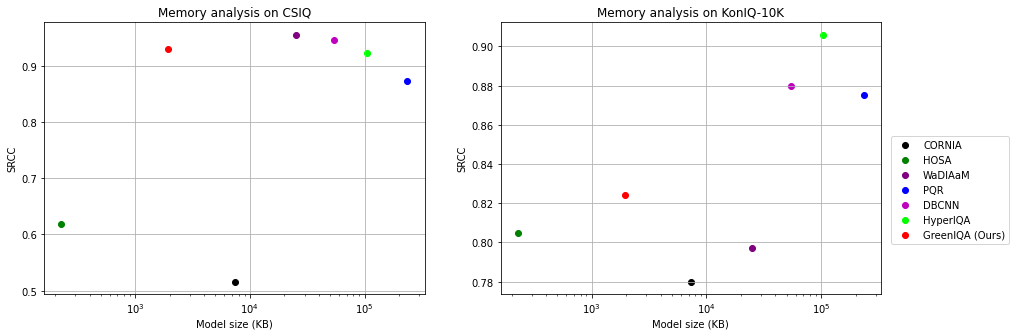}\\
(a) SROCC performance versus the memory size of several benchmarking methods  \\
\includegraphics[width=1\linewidth]{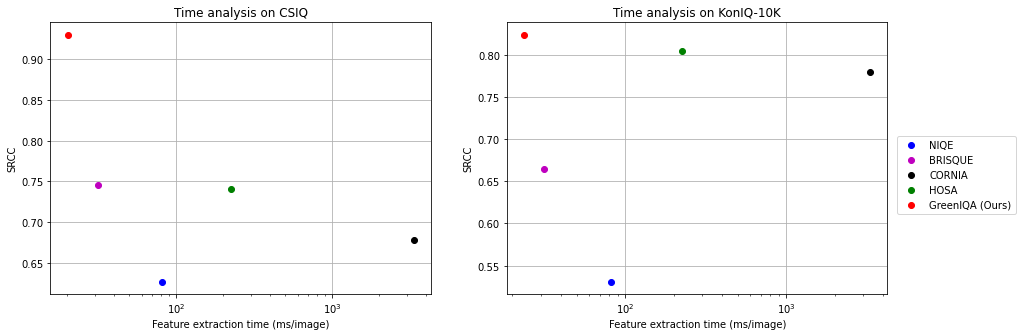}\\ 
(b) SROCC performance versus the running time of several benchmarking methods
\caption{Comparison of (a) model sizes and (b) running time 
on CSIQ and KonIQ-10K.}\label{fig:analysis}
\end{figure*}


\subsection{Memory Size Comparison} A small memory size is critical to
real-world applications such as mobile phones.  To illustrate the
tradeoff between performance and memory sizes, we show the SROCC
performance versus the memory size (under the assumption that one
parameter takes one byte memory) of several benchmarking methods against
CSIQ and KonIQ-10K in Fig. \ref{fig:analysis}(a). We see that GreenBIQA
can achieve simliar performance on CSIQ and competitive performance on
KonIQ-10K. Since high performance deep-learning methods have a huge
pre-trained neural network as their backbone, their model sizes are
usually more than 100 MB. In contrast, GreenBIQA only needs 600 KB for
feature extraction and 1.17 MB for the regression model. Furthermore,
GreenBIQA outperforms BIECON and WaDIQaM, which are two deep-learning
methods without pre-training, on KonIQ-10K. As compared to
codebook-learning methods such as CORNIA and HOSA, whose model sizes are
determined by the lengths of the codewords, GreenBIQA achieves much
better performance while its size is slightly larger than HOSA and much
smaller than CORNIA. 

\subsection{Running Time Comparison} 
Another factor to consider for BIQA at
the client is running time in MOS prediction. As streaming services and
video calls are widely used at mobile devices nowadays, it is desired to
assess the quality of individual video frames in real-time with limited
computation power. For deep-learning methods with huge pre-trained
models, their computations are heavy and have to be run on one or multiple
GPUs. To illustrate the tradeoff between performance and running time,
we show the SROCC performance versus the MOS inference time per frame of
GreenBIQA and four conventional methods on a CPU against CSIQ and
KonIQ-10K in Fig.  \ref{fig:analysis}(b). Note that we do not include
deep-learning-based methods in Fig.  \ref{fig:analysis}(b) since they
are all implemented on GPUs. GreenBIQA can process around 43 images per
second with a single CPU. Thus, it meets the real-time processing
requirement. We see from Fig.  \ref{fig:analysis}(b) that GreenBIQA has
the smallest inference time among all benchmarking methods. It can be
executed faster than conventional methods (i.e. NSS-based and
codebook-learning methods) with significantly better SROCC performance on
both synthetic- and authentic-distortion datasets.

\section{Conclusion and Future Work}\label{sec:conclusion}

A lightweight BIQA method, called GreenBIQA, was proposed in this
paper.  GreenBIQA was evaluated on two synthetic and two authentic
datasets.  GreenBIQA outperforms all conventional BIQA methods in all
datasets. As compared to pre-trained deep-learning methods, GreenBIQA
achieves competitive performance with 54x smaller model size. It can
extract quality-aware features in real-time (i.e. 43 images per second)
using only CPUs. Generally speaking, human annotations of image quality
are often limited and difficult to collect.  Thus, it is desired to
transfer the GreenBIAQ model trained by one model to another one with
weak supervision in the future extension. 

\bibliographystyle{unsrt}  
\bibliography{references}

\end{document}